\documentclass{ws-ijmpc}

\begin{document}

\markboth{F. L. Dubeibe}
{Solving the time--dependent Schr\"{o}dinger equation in Mathematica 6.0}

\catchline{}{}{}{}{}

\title{SOLVING THE TIME--DEPENDENT SCHR\"{O}DINGER EQUATION WITH ABSORBING BOUNDARY CONDITIONS AND SOURCE TERMS IN MATHEMATICA 6.0}

\author{F. L. DUBEIBE}

\address{{\it Facultad de Ciencias Humanas y de la Educaci\'on, Escuela de Pedagog\'ia y Bellas Artes, \\
Universidad de los Llanos, Villavicencio, Colombia}\\
{\it fldubeibem@unal.edu.co}}

\maketitle

\begin{history}
\received{13 September 2010}
\revised{04 October 2010}
\end{history}

\begin{abstract}
In recent decades a lot of research has been done on the numerical solution of the time--dependent Schr\"{o}dinger equation. On the one hand, some of the proposed numerical methods do not need any kind of matrix inversion, but source terms cannot be easily implemented into this schemes; on the other, some methods involving matrix inversion can implement source terms in a natural way, but are not easy to implement into some computational software programs widely used by non-experts in programming ({\it e.g.} Mathematica). We present a simple method to solve the time--dependent Schr\"{o}dinger equation by using a standard Crank-Nicholson method together with a Cayley's form for the finite-difference representation of evolution operator.  Here, such standard numerical scheme has been simplified by inverting analytically the matrix of the evolution operator in position representation. The analytical inversion of the $N\times N$ matrix let us easily and fully implement the numerical method, with or without source terms, into Mathematica or even into any numerical computing language or computational software used for scientific computing.

\end{abstract}

\ccode{PACS Nos.: 01.50.H-, 02.70.Bf, 02.10.Ud, 02.60.Cb, 03.65.Ge}

\section{Introduction}
One of the main arguments used to explain why quantum mechanics is not easily accessible to most of the students attending for first time a quantum mechanics course, is that the physical situations of quantum mechanics are not everyday life phenomena and that can be approached only through abstract mathematics. The significant differences between the classical and quantum physics, makes difficult to our macroscopically-trained minds to imagine what is happening in a physical situation at the quantum regime. On the other hand, usually in the classical mechanics courses as a general problem-solving strategy is suggested first to draw a sketch or diagram that represents the physics of the problem under consideration.\cite{01,02} Such strategy is contradicted in the case of quantum mechanical problems where in almost any introductory quantum mechanics textbook can be found statements as: do not try to imagine the physical situation.\cite{03,04} Told to avoid visualization, the students fall into misunderstandings because the lack of a mental picture leads to inefficient problem solving \cite{05}.

From the previous discussion and as has been considered for some other authors,\cite{06,07} it is necessary to implement visualization techniques that can improve both, understanding of, and problem-solving in quantum mechanics. In the present paper we concentrate particularly in particle propagation methods. Along this line, since the pioneering work by Feit {\it et al.},\cite{7a} some efforts has been done aiming for a comprehensive and easy implementation (even for beginners in the field) of numerical methods to solve the Schr\"{o}dinger equation (see for instance  Ref.~\refcite{7b} and references therein).  Yet, it should be noted that nowadays there are many programs that are able to do such simulations,\cite{08,09,10} but in most cases, those are numerical codes that require more than a basic knowledge of programming to be implemented. 

Writing a simple program that can be fully implemented by the student allows a better understanding of the phenomenon, opens the possibility of treating a wider range of problems and allows the student to have a pleasant first contact with programming. In this paper we describe a numerical integration method for the time--dependent Schr\"{o}dinger equation and the implementation of absorbing boundary conditions and source terms into this scheme. Unlike most existing methods,\cite{11,12} this method is easy to implement, versatile, very accurate, and can be implemented without including any kind of matrix inversion package, which allows a fully implementation on a mathematical package such as Mathematica, providing easily visualizable results of the evolving system.

The plan of this paper is as follows. In section \ref{themethod} we present a short description of the technique and its numerical implementation. We follow by presenting the formula for the inverse of a non-symmetrical tridiagonal Jacobian matrix and we show how to introduce the inverse matrix formulas into the numerical scheme. In section \ref{metabc} we continue relating it to the absorbing boundary conditions to finally apply it for the case of the source term in section \ref{metst}. We end this paper by comparing our numerical results with the analytical solution for the Schr\"{o}dinger equation with source term to finally compare the analytical and numerical results for the transmission probability of a finite potential barrier.

\section{The Method}\label{themethod}
Let us consider the Schr\"{o}dinger equation in atomic units, i.e. $m=\hbar=1$,
\begin{equation}\label{eq1c2}
i \frac{\partial}{\partial t}\psi(x,t)=H(x,t)\psi(x,t),
\end{equation}
with the Hamiltonian given by
\begin{equation}\label{eq2c2}
H(x,t)=-\frac{1}{2}\frac{\partial^2}{\partial x^2}+V(x,t).
\end{equation}
The idea is to compute the time evolution of the wave function $\psi(x,t)$ for $t>t_0$, given an initial state $\psi(x,t_0)$. We start by dividing the time interval into $n$ subintervals of equal length $\Delta t=(t-t_0)/n$, and use an implicit Crank-Nicholson integrator scheme \cite{13} to propagate the wave function from one time step to the next one.

The formal solution to Eq. (\ref{eq1c2}) could be expressed in terms of the time evolution operator as,
\begin{equation}
\psi(x,t)=e^{-i H t}\psi(x,0).
\end{equation}
The effective time evolution operator $\mathcal{U}$ for one discrete
time step $\Delta t$, can be expressed using Cayley's form for the finite-difference representation of $e^{-i H t}$, which is a combination of a fully implicit and a fully explicit method,\cite{14}
\begin{equation}\label{evol}
\mathcal{U}(t+\Delta t, t)= \frac{1-\frac{i\Delta t}{2}H(x,t)}{1+\frac{i\Delta t}{2}H(x,t)}.
\end{equation}
Such representation of $\mathcal{U}$ is second-order accurate in space and time and also unitary. The integration scheme for the wave function then reads
\begin{equation}\label{intsch}
\left(1+\frac{i\Delta t}{2}H(x,t)\right)\psi(x,t+\Delta
t)=\left(1-\frac{i\Delta t}{2}H(x,t)\right)\psi(x,t).
\end{equation}
The wave function can be expanded on a discrete lattice as
\begin{equation}
\psi(x,t_n)=\sum_{j=1}^{N} \psi_{j}^{n}\chi_{j}\,,
\end{equation}
where $\psi_{j}^{n}=\psi(x_{j}, t_{n})$ is the value of the wave
function at the position $x_{j}$ of the $j$th lattice site at time
$t_{n}=t_0 + n\Delta t$, with a grid basis
\begin{equation}\label{grid}
\chi_{j}=\left\{\begin{array}{ll}
                  1, & x_{j}-\frac{1}{2}\Delta x \leq x \leq x_{j} + \frac{1}{2}\Delta x ; \\
                  0, & {\rm{otherwise}}.
\end{array}\right.
\end{equation}
Here $\Delta x=(x_{\rm max}- x_{\rm min})/N$, with $x_{\rm max}$ and $x_{\rm min}$
the boundaries of the finite grid.

Using the finite-difference representation
for the kinetic part of the hamiltonian,\cite{15} we have
\begin{equation}
\left(1\pm\frac{i\Delta t}{2}H\right)\psi(x_j,t_n) \simeq
\psi_{j}^{n} \pm \frac{i\Delta t}{2}\left(- \frac{\psi_{j+1}^{n}- 2
\psi_{j}^{n}+ \psi_{j-1}^{n}}{2 \Delta x^{2}}+ V_{j}^{n}
\psi_{j}^{n} \right)
\end{equation}
with $V_{j}^{n}=V(x_{j}, t_{n})$. By introducing
$\vec{\psi}^{n}=(\psi_{1}^{n},..., \psi_{j}^{n},..., \psi_{N}^{n} )$,
the lattice representation of Eq. (\ref{intsch}) finally reads
\begin{equation}\label{method}
\vec{\psi}^{n+1}= \mathbf{D}_{2}^{-1}\mathbf{D}_{1}
\vec{\psi}^{n}\,,
\end{equation}
where we define
\begin{equation}
\mathbf{D}_{1}= \left(1 - \frac{i\Delta
t}{2}H\right)=(1-\mathbf{S}),\quad \mathbf{D}_{2} = \left(1 +
\frac{i\Delta t}{2}H\right)=(1+\mathbf{S}),
\end{equation}
with $\mathbf{S}=\frac{i\Delta t}{2}H$. The matrix product can be rewritten as
\begin{equation}
\mathbf{D}_{2}^{-1}\mathbf{D}_{1} = (1+\mathbf{S})^{-1}(1-\mathbf{S}) = 2\mathbf{D}_{2}^{-1} -1,\label{D2s}
\end{equation}
then, the wave packet evolution is achieved just by
inverting the matrix $\mathbf{D}_{2}$.

For the case of time independent potentials, the explicit
$N\times N$ representation of $\mathbf{D}_{1}$ and $\mathbf{D}_{2}$
reads

\begin{equation}\label{d1}
\mathbf{D}_{1}= \left(
  \begin{array}{cccccc}
\gamma_{1}& \alpha &      &       &       &  \\
\alpha & \gamma_{2}& \alpha  &       &       &  \\
        & \alpha & \gamma_{3} & \alpha  &       &  \\
        &     &\ddots&\ddots &\ddots &  \\
        &     &      &\alpha&\gamma_{N-1}& \alpha \\
        &     &      &       &\alpha&\gamma_{N} \\
  \end{array}
\right)
,
\mathbf{D}_{2}= \left(
  \begin{array}{cccccc}
\xi_{1}& -\alpha &      &       &       &  \\
- \alpha & \xi_{2}& -\alpha  &       &       &  \\
        & -\alpha & \xi_{3}  &  -\alpha  &       &  \\
        &     &\ddots&\ddots &\ddots &  \\
        &     &      &- \alpha&\xi_{N-1}& -\alpha \\
        &     &      &       &- \alpha& \xi_{N} \\
  \end{array}
\right)
\end{equation}
with
\begin{equation}\label{param_tid}
\alpha=\frac{i \Delta t}{4 \Delta x^{2}}, \,\,\,
\gamma_{j}=1-\beta_{j}, \,\,\, \xi_{j}=1+\beta_{j}, \,\,\, {\rm{and}}
\,\,\, \beta_{j}=\frac{i\Delta t}{2}\left(\frac{1}{\Delta x^{2}}+
V_{j} \right).
\end{equation}

\subsection{Inverse of a Tridiagonal Matrix}
Let us consider the $N\times N$ nonsingular tridiagonal matrix
$\mathbf{D}$
\begin{equation}\label{D}
\mathbf{D}= \left(
  \begin{array}{cccccc}
    a_1 & b_1 &      &       &       &  \\
    c_1 & a_2 & b_2  &       &       &  \\
        & c_2 & a_3  &  b_3  &       &  \\
        &     &\ddots&\ddots &\ddots &  \\
        &     &      &c_{N-2}&a_{N-1}& b_{N-1} \\
        &     &      &       &c_{N-1}& a_{N} \\
  \end{array}
\right)
\end{equation}

Usmani \cite{16} gave an elegant and concise formula for the
inverse of the tridiagonal matrix\footnote{A few typos and misprints from the original paper were corrected.}: 

\begin{equation}\label{invanal}
(\mathbf{D})^{-1}_{ij}=\left\{\begin{array}{ll}
                  (-1)^{i+j} b_i\ldots b_{j-1}\theta_{i-1}\phi_{j+1}/\theta_{N}, & i\leq j; \\
                  (-1)^{i+j} c_j\ldots c_{i-1}\theta_{j-1}\phi_{i+1}/\theta_{N}, & i> j.
\end{array}\right.
\end{equation}
where $\theta_{i}$ satisfy the recurrence relation
\begin{equation}
\theta_i = a_i \theta_{i-1} - b_{i-1} c_{i-1} \theta_{i-2}, \qquad
{\rm{for}} \qquad i = 2,\ldots, N,
\end{equation}
with initial conditions $\theta_0 = 1$ and $\theta_1 = a_1$, and
$\phi_{i}$ satisfy the recurrence relation
\begin{equation}
\phi_i = a_i \phi_{i+1} - b_{i}c_{i}\phi_{i+2}, \qquad {\rm{for}}
\qquad i = N-1,\ldots, 1,
\end{equation}
with initial conditions $\phi_{N+1} = 1$, $\phi_N = a_N$, and
$\theta_N = {\rm{det}}\,\mathbf{D}$.

Using the last procedure, we obtain a simplified formula for the inverse of the matrix $\mathbf{D}_{2}$,
\begin{eqnarray}\label{inverse}
(\mathbf{D}_{2})^{-1}_{ij}=d_{ij}=(-1)^{i+j}
(-\alpha)^{|j-i|}\theta_{i-1}\phi_{j+1}/\theta_{N},\quad
{\rm{if}}\quad i\leq j\,.
\end{eqnarray}
Due to the tridiagonal symmetric nature of $\mathbf{D}_{2}$, the inverse satisfies, $d_{ji}=d_{ij}$. The recurrence relations are given by
\begin{equation}
\theta_i = \xi_i \theta_{i-1} - \alpha^2 \theta_{i-2}, \qquad
{\rm{for}} \qquad i = 2,\ldots, N,
\end{equation}
and
\begin{equation}
\phi_i = \xi_i \phi_{i+1} - \alpha^2 \phi_{i+2}, \qquad {\rm{for}}
\qquad i = N-1,\ldots, 1,
\end{equation}
with $\theta_0 = 1$, $\theta_1 = a_1$,  $\phi_{N+1} = 1$, and $\phi_N = a_N$.

Finally, the elements of the matrix product
$\mathbf{E}=\mathbf{D}_{2}^{-1}\mathbf{D}_1=2\mathbf{D}_{2}^{-1} -
1$, are given by
\begin{equation}\label{matrixE}
(\mathbf{E})_{ij}= 2\,d_{i j} - \delta_{ij}\quad {\rm{for}}\quad
i,j=1,\ldots, N\,,
\end{equation}
where $\delta_{ij}$ is the Kronecker delta.

In what follows we consider two standard systems belonging to the class of time independent potentials:
The finite square potential well and the finite square barrier. In both cases
Dirichlet boundary conditions are assumed. These boundary conditions may
cause unwanted reflections, therefore, we have to perform the numerical
calculations on a sufficiently large bounded interval, placing the impinging particle far
away of the numerical boundary and restricting the time interval such that the
reflections do not affect the solution in the region of interest.

\subsection{Finite Square Potential Well and Finite Square Barrier}
\begin{figure}[h!]
\begin{center}
\includegraphics[width=7cm,angle=270]{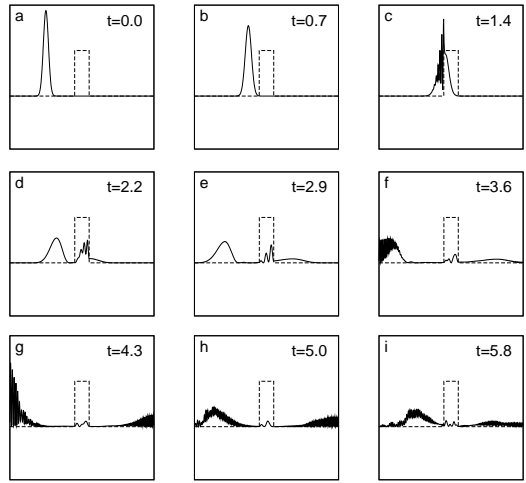}
\caption{Gaussian wave-packet scattering from a finite square potential barrier. The initial conditions
are $x_0=-10, \sigma_{0}=1, x_{\rm b}=2$, $t_{0}=0$ and $p_0=7$. The left and right borders of
the domain are $x_{\rm min} = -20$ and $x_{\rm max} = 20$ respectively. The
size of every lattice in the grid is $\Delta x = 0.04$ with $N=1000$
discrete lattices in all the spatial domain and the time step is
$\Delta t = 0.002$. With the given parameters the matrix elements
are calculated from Eq. (\ref{param_tid}). The parameter $t=t_{0}+ n \Delta t$ denotes the time of each configuration.} \label{evpot+}
\end{center}
\end{figure}

\begin{figure}[h!]
\begin{center}
\includegraphics[width=7cm,angle=270]{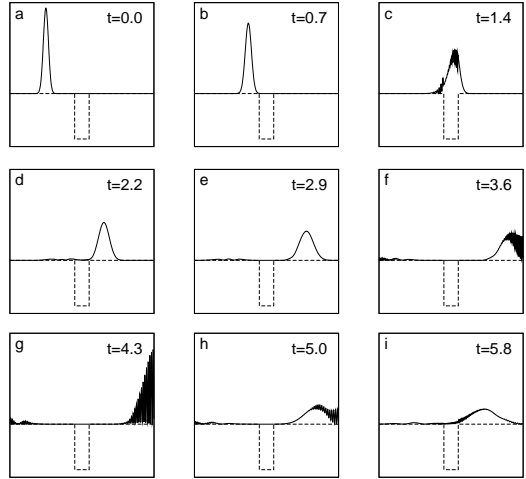}
\caption{Gaussian wave-packet scattering from a finite square potential well. The same parameters as
in Fig. \ref{evpot+}. } \label{evpot-}
\end{center}
\end{figure}

In order to test the method and to observe the effect of the
Dirichlet boundary conditions, we consider a Gaussian wave packet
\begin{equation}\label{Iniwp}
\psi(x,0)=\sqrt[4]{\frac{1}{\sigma_{0}^{2}\pi}}\exp\left[i p_0 (x-x_{0}) -\frac{(x-x_0)^{2}}{2\sigma_{0}^{2}}\right],
\end{equation}
initially centered at $x_0$, with average
momentum $p_0$ and initial width $\sigma_{0}$, which moves into the region of a short range potential defined as
\begin{eqnarray}\label{pot}
V(x)= \left\{
  \begin{array}{ll}
    \pm p_{0}^{2}/2, & \hbox{$-x_{\rm b}<x<x_{\rm b}$;} \\
    0, & \hbox{otherwise.}
  \end{array}
\right.
\end{eqnarray}
This is a potential barrier (plus signed) or a potential well (minus signed) whose height or depth, respectively, equals the average energy
$p_{0}^2 /2$ of the Gaussian wave packet.\cite{17} In Fig. \ref{evpot+} and Fig. \ref{evpot-} we show the Gaussian wave packet scattering from
the finite square barrier and the finite square well respectively.

As can be seen in Figs. \ref{evpot+} and \ref{evpot-}, there exists a strong back reflection effect due to
the Dirichlet boundary conditions in both cases. The wave packet behaves like
inside a large infinite square well of length $L=2 x_{\rm max}$. If the wave packet spreads quickly, any reflected portion of the wave
will then interfere with the portion of the incident wave, giving rise to a
non-physical interference pattern. This situation imposes limitations on the choice of the input
parameters, e.g. $x_0$ and $\sigma_{0}$ in Eq. (\ref{Iniwp}) must be
chosen so that $\psi(-L/2,0)$ and $\psi(L/2,0)$ are essentially zero
at least at the beginning $t=t_0$.

For the square barrier case, Fig. \ref{evpot+}, a fraction of the wave packet is captured by the barrier
and remains trapped for a period which is longer than the time of
transmission through the barrier. The captured piece of wave packet
bounces back and forth between the barrier walls with a small amount
of probability escaping in each collision till the entire packet
escapes. With the present numerical scheme the dynamic evolution of the trapped wave-packet can be easily observed at each time step.

\section{Method with Absorbing Boundary Conditions (ABC)}\label{metabc}
The numerical solutions of the time--dependent Schr\"{o}dinger
equation provide us insight into the dynamics of quantum mechanical
systems. However, as was discussed in the previous section in
practical calculations the area of computation must be limited to a
finite grid because of the finite capacity of the computer memories. 
This finite grid produces undesirable reflections at the
artificial boundaries of the area of computation. To minimize this
artificial effect we implement in the present section the
so called absorbing boundary conditions. These are local boundary
conditions that approximate the one way wave equation of a wave
function.

Let us set the following equation as the starting point of the
discussion
\begin{equation}\label{schrtind}
i \hbar\frac{\partial}{\partial
t}\psi(x,t)=\left(-\frac{\hbar^{2}}{2
m}\frac{\partial^2}{\partial x^2}+V(x)\right)\psi(x,t).
\end{equation}
In order to obtain the formulas for the absorbing boundary
conditions, following Shibata,\cite{18} we consider the special solutions
$\psi(x,t)=\exp(-i(\omega t - k x))$, these are states of definite
energy $E$ satisfying the dispersion relation,
\begin{equation}\label{disp2}
\hbar k=\pm \sqrt{2m(\hbar \omega - V)}.
\end{equation}

The absorbing boundary conditions must be designed to satisfy the dispersion relation given
by the plus signed Eq. (\ref{disp2}) at the boundary $x_{\rm max}$ and
the minus signed at the boundary $x_{\rm min}$. However, function
(\ref{disp2}) is not rational and cannot be converted into a partial
differential equation, nonetheless, this relation can be linearly approximated by
\begin{equation}\label{disp_ap}
\hbar k=g_{1}(\hbar \omega - V)+g_{2}\,,
\end{equation}
with
\begin{equation}\label{g1g2}
g_{1}=\pm\frac{\sqrt{2 m\alpha_{2}}-\sqrt{2
m\alpha_{1}}}{\alpha_2 - \alpha_1},\quad
g_{2}=\pm\frac{\alpha_{2} \sqrt{2 m\alpha_{1}}-\alpha_{1}\sqrt{2
m\alpha_{2}}}{\alpha_2 - \alpha_1}\,.
\end{equation}
The correspondence of $\partial/\partial
t\Leftrightarrow -i \omega$ and $\partial/\partial x\Leftrightarrow
i k$ leads us to rewrite Eq. (\ref{g1g2}) into the partial
differential equation
\begin{equation}\label{newpd}
i \hbar\frac{\partial}{\partial t}\psi(x,t)=\left(-i
\hbar\frac{1}{g_{1}\partial x}+ V
-\frac{g_{2}}{g_{1}}\right)\psi(x,t).
\end{equation}

Now we outline how to incorporate the ABC into the lattice
representation of the wave function (with $\hbar=m=1$). The idea
is to replace the differential equation for the boundary components
$\psi_{N}^{n}$ and $\psi_{1}^{n}$ of the state vector
$\vec{\psi}^{n}$. As was discussed by Paul {\it et al.},\cite{19} in order to
obtain an accurate expression for the derivative at the borders of
the grid is convenient to introduce an intermediate point $\bar{x}$
between the last two points of each side of the grid, then for
example, at the right hand side the wave function must be replaced by
\begin{equation}
\psi(\bar{x},t)\simeq \frac{1}{2}[\psi(x_{N},t)+\psi(x_{N-1},t)].
\end{equation}
In the grid representation, the finite-difference equation for the
right and left sides reads
\begin{eqnarray}
\frac{i}{2 \Delta t}(\psi_{N}^{n+1}+
\psi_{N-1}^{n+1}-\psi_{N}^{n}-\psi_{N-1}^{n})&=&\frac{-i}{g_{1}\Delta
x}(\psi_{N}^{n}-\psi_{N-1}^{n})\nonumber
\\
&+&
\frac{1}{2}\left(V-\frac{g_2}{g_1}\right)(\psi_{N}^{n}+\psi_{N-1}^{n})\,,
\label{abcgrid1}
\end{eqnarray}
and
\begin{eqnarray}
\frac{i}{2 \Delta t}(\psi_{2}^{n+1}+
\psi_{1}^{n+1}-\psi_{2}^{n}-\psi_{1}^{n})&=&\frac{i}{g_{1}\Delta
x}(\psi_{2}^{n}-\psi_{1}^{n})
\nonumber\\
&+&
\frac{1}{2}\left(V-\frac{g_2}{g_1}\right)(\psi_{2}^{n}+\psi_{1}^{n})\,,\label{abcgrid2}
\end{eqnarray}
respectively. The equations (\ref{abcgrid1}) and (\ref{abcgrid2}) allow a straightforward
incorporation into the matrix representation.

The new matrices $\mathbf{D}_{1,2}$ are given by
\begin{equation}\label{d1n}
\hspace{-2.5cm}\mathbf{D}_{1}= \left(
  \begin{array}{cccccc}
\eta_{4}& \eta_{3} &      &       &       &  \\
\alpha & \gamma_{2}& \alpha  &       &       &  \\
        & \alpha & \gamma_{3}  & \alpha  &       &  \\
        &     &\ddots&\ddots &\ddots &  \\
        &     &      &\alpha&\gamma_{N-1}& \alpha \\
        &     &      &       &\eta_{3} &\eta_{4}  \\
  \end{array}
\right),
\mathbf{D}_{2}= \left(
  \begin{array}{cccccc}
\eta_{2}& \eta_{1} &      &       &       &  \\
- \alpha & \xi_{2}& -\alpha  &       &       &  \\
        & -\alpha & \xi_{3}  &  -\alpha  &       &  \\
        &     &\ddots&\ddots &\ddots &  \\
        &     &      &- \alpha&\xi_{N-1}& -\alpha \\
        &     &      &       &\eta_{1}& \eta_{2} \\
  \end{array}
\right)
\end{equation}
with
\begin{eqnarray}
&&\eta_{1} \equiv \eta_{2} \equiv \frac{i}{2 \Delta t}\,,
\\
&&\eta_{3} \equiv \frac{i}{2 \Delta t} + \frac{i}{g_{1} \Delta x} +
\frac{1}{2} \left(V- \frac{g_{2}}{g_{1}}\right)\,,
\\
&&\eta_{4} \equiv \frac{i}{2 \Delta t} - \frac{i}{g_{1} \Delta x} +
\frac{1}{2} \left(V- \frac{g_{2}}{g_{1}}\right)\,.
\end{eqnarray}

The main cause of artificial back reflection for plane waves in the
presence of the above boundary conditions comes from the approximate
nature of the finite difference evaluation. Clearly, these
approximations become better decreasing the grid spacing
$\Delta x$.

In the present case, the matrix elements in (\ref{D}) are given by
$a_{1}=a_{N}=\eta_{2},$ $b_{1}=c_{N-1}=\eta_{1},$  $b_{j}=-\alpha$
for $j=2,...,N-1$ and $c_{j}=-\alpha$ for $j=1,...,N-2$. The
elements in $(\mathbf{D}_{2})^{-1}_{ij}=d_{ij}$ are given by
\begin{eqnarray*}
&&d_{11}=\phi_2/\theta_N, \\
&&d_{1j}=(-1)^{1+j} \eta_2 (-\alpha )^{|j-2|}
\phi_{j+1}/\theta_N, \quad{\rm{for}}\quad{j=2,...,N} \\
&&d_{ij}=(-1)^{i+j}(-\alpha)^{|j-i|}\theta_{i-1}\phi_{j+1}/\theta_{N},
\quad{\rm{for}}\quad{i,j=2,...,N}\quad{\rm{with}}\quad{i\leq j} \\
&&d_{ij}=(-1)^{i+j}(-\alpha)^{|i-j|}\theta_{j-1}\phi_{i+1}/\theta_{N},
\,\,\,{\rm{for}}\,\,\,{i=2,...,N-1,\,\,\,j=1,...,N}\,\,\,{\rm{with}}\,\,\,{i > j} \\
&&d_{Nj}=(-1)^{N+j}\eta_{1}(-\alpha)^{|N-j-1|}\theta_{j-1}/\theta_{N}, \quad{\rm{for}}\quad{j=1,...,N-1}.
\end{eqnarray*}
with
\begin{eqnarray*}
&&\theta_0 = 1 , \\
&&\theta_{1} = \eta_{2} , \\
&&\theta_{2} = \xi_{2} \eta_{2} + \alpha \eta_{1} , \\
&&\theta_{i} =\xi_{i}\theta_{i-1}- \alpha ^2 \theta_{i-2}, \quad{\rm{for}}\quad{i=3,...,N-1} \\
&&\theta_{N} = \eta_{2}\theta_{N-1} + \alpha  \eta_{1} \theta_{N-2}
\end{eqnarray*}
and
\begin{eqnarray*}
&&\phi_{N+1} = 1, \\
&&\phi_{N} = \eta_{2}, \\
&&\phi_{N-1} = \xi_{N-1}\eta_{2} + \alpha \eta_{1}, \\
&&\phi_{i} = \xi_{i}\phi_{i+1} - \alpha^2 \phi_{i+2},\quad{\rm{for}}\quad{i=N-2,...,2} \\
&&\phi_{1} = \eta_{2}\phi_{2} + \alpha \eta_{1}\phi_{3}.
\end{eqnarray*}
On the other hand, the non-symmetric character of the matrices
$\mathbf{D}_{1,2}$ does not let us to write the matrix product
$\mathbf{E}=\mathbf{D}_{2}^{-1} \mathbf{D}_{1}$ as simple as in Eq.
(\ref{D2s}). The new components of the product are
\begin{eqnarray*}
&&E_{i1} = \eta_{4} d_{i1}+ \alpha d_{i2},
\quad{\rm{for}}\quad{i=1,...,N} \\
&&E_{i2} = \eta_{3} d_{i1}+ \gamma_{2} d_{i2}+ \alpha d_{i3},
\quad{\rm{for}}\quad{i=1,...,N} \\
&&E_{ij}=\alpha  d_{i j-1}+ \gamma_{j} d_{ij} + \alpha d_{i j+1}
\quad{\rm{for}}\quad{i=1,...,N}, {j=3,...,N-2} \\
&&E_{i N-1} = \alpha  d_{i N-2} + \gamma_{N-1} d_{i N-1} + \eta_{3}
d_{i N} \quad{\rm{for}}\quad{i=1,...,N} \\
&&E_{i N} = \alpha  d_{i N-1} + \eta_{4} d_{i N} \quad{\rm{for}}\quad{i=1,...,N}
\end{eqnarray*}
It should be noted that the parameters in Eq. (\ref{param_tid}) are the same, but now these parameters are defined between $2$ and $N-1$.

\subsection{Finite Square Potential Well and Finite Square Barrier with ABC}
\begin{figure}[h!]
\begin{center}
\includegraphics[width=7cm,angle=270]{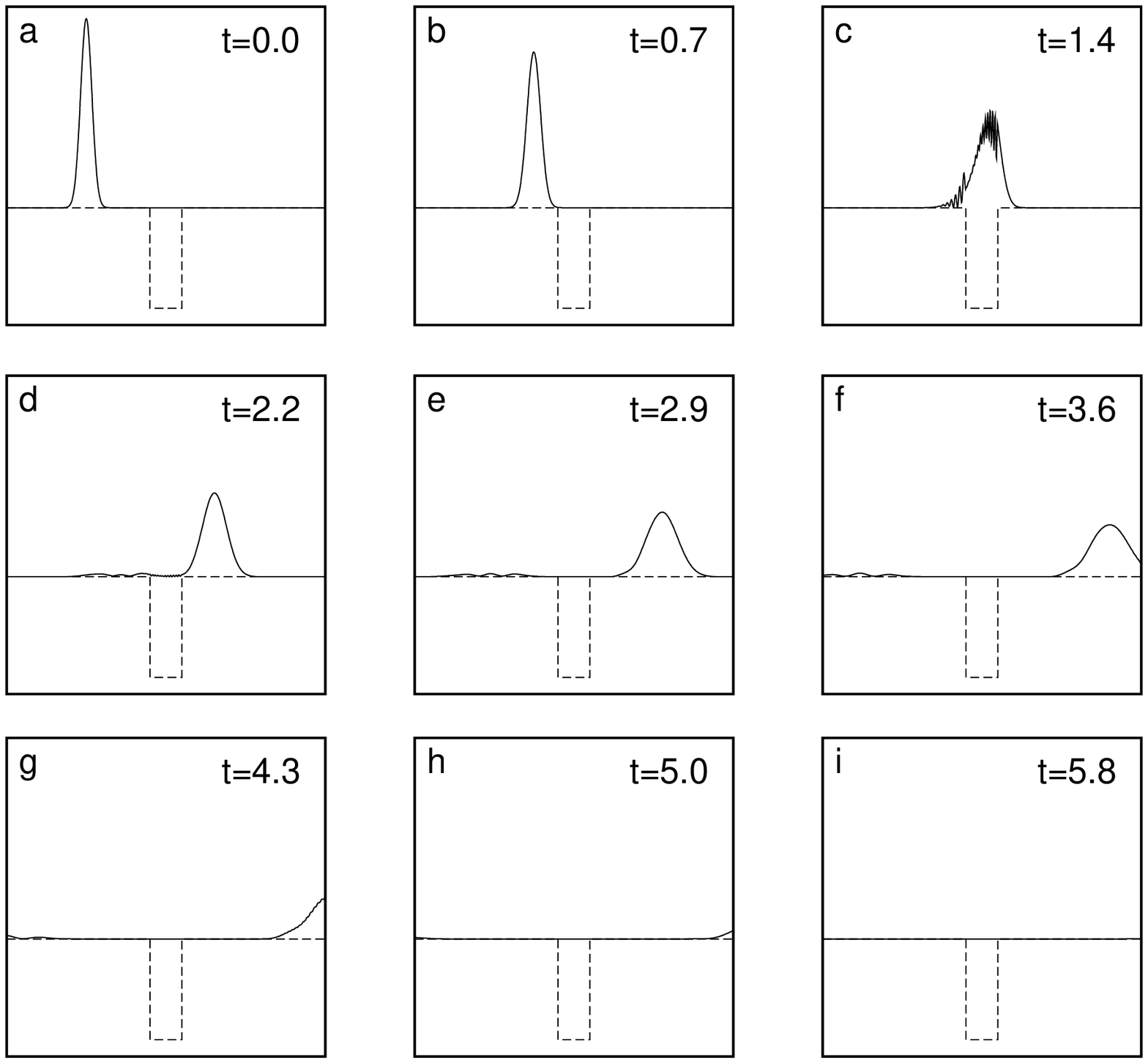}
\caption{Gaussian wave-packet scattering from a finite square potential well in the presence of ABC.
In this case we choose $\alpha_1=24$ and $\alpha_2=25$, the other
initial conditions and parameters are the same as in Fig. \ref{evpot-}} \label{evpot-abc}
\end{center}
\end{figure}

\begin{figure}[h!]
\begin{center}
\includegraphics[width=7cm,angle=270]{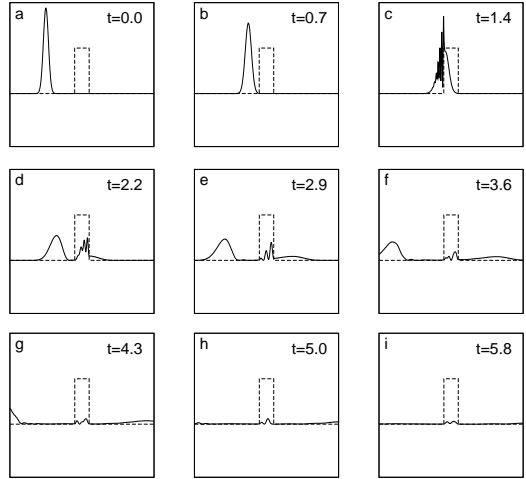}
\caption{Gaussian wave-packet scattering from a finite square potential barrier in the presence of ABC.
The same parameters as in Fig. \ref{evpot+}.} \label{evpot+abc}
\end{center}
\end{figure}

In order to observe the effect of the ABC at the boundaries of the grid,
we use the same examples as before: The square well and the square barrier with the Gaussian wave
packet Eq. (\ref{Iniwp}) as test particle. The potential function is defined as in Eq. (\ref{pot}). In Fig. \ref{evpot-abc} and Fig. \ref{evpot+abc}, we show a wave packet scattering off a square well and a square barrier respectively.

As can be seen from the Figs. \ref{evpot-abc} and \ref{evpot+abc}, the boundary conditions effectively reduce the non-physical reflections of the impinging wave packet at the boundary of the computation area. The clear presentation of the resonances let us study its dynamical evolution in a detailed way.

\section{Method with Absorbing Boundary Conditions and Source Term}\label{metst}
As a last example illustrating the validity and effectiveness of our
simplified method, we consider the presence of a source term. The
equation of motion now reads
\begin{equation}\label{schrST}
i \frac{\partial}{\partial
t}\psi(x,t)=H(x,t)\psi(x,t)+S(t)\exp(-i\omega t)\delta(x)\,,
\end{equation}
where $S(t)=S_0 [1-\exp(-t/\Delta T)]$. This function provides a
smooth evolution of the source term towards the desired final value
$S(t\rightarrow \infty)=S_0$.  In order to have an analytical result, we consider the stationary
solutions of Eq. (\ref{schrST}) for the particular case $V(x)=0$ with
$S(t)=S_0$. Introducing the Fourier transformed wave function
$\tilde{\psi}(q,t)=\int \exp(iqx)\psi(x,t)dq$ and considering the
previous conditions, the Eq. (\ref{schrST}) takes the form
\begin{equation}
\left(i\frac{\partial}{\partial t}-\frac{\hbar^{2}q^{2}}{2
m}\right)\tilde{\psi}(q,t)=S_0 \exp(-i\omega t).
\end{equation}
This equation admits solutions of the form
$$\tilde{\psi}(q,t)= \frac{2 S_0}{k^{2}-q^{2}}\exp(-i\omega t)\,, $$
where $k^{2}=2\omega$. Transforming back to the configuration space,
the solution is given by
\begin{equation}\label{solv0}
\psi(x,t)=\frac{S_0}{ik}\exp(ik|x|)\exp(-i\omega t),
\end{equation}
showing that the source emits in both direction a monochromatic
wave Eq. (\ref{solv0}).

As it was shown by Paul {\textit{et. al.}},\cite{19} working with a grid representation
of the wave function, it is convenient to approximate the $\delta$
function by
\begin{equation}
R(x)=\frac{1}{\Delta x}[\Theta(x+\Delta x/2)-\Theta(x-\Delta x/2)],
\end{equation}
where $\Theta$ is the Heaviside step function. With this
approximation, the error scales quadratically with the size of the
grid $\Delta x$ and becomes negligible for reasonable small values
of $\Delta x$. The implementation of the source term at position
$x_{j'}$ in the grid representation reads as
$$S_{j}^{n}=S(t_{n})\exp(-i\omega t_{n})\delta_{j,j'}$$
where $\delta_{j,j'}=1$ if $j=j'$ and $0$ otherwise. In the presence
of the source term, Eq. (\ref{method}) is given by
\begin{equation}\label{methodST}
\vec{\psi}^{n+1}= \mathbf{D}_{2}^{-1}(\mathbf{D}_{1} \vec{\psi}^{n}-\vec{b}^{n}),
\end{equation}
where the components of $\vec{b}^{n}$ are defined as
\begin{equation}
b_{j}^{n}=\frac{i\Delta t}{2}(S_{j'}^{n}+S_{j'}^{n+1})\delta_{j,j'}.
\end{equation}
The numerical implementation is the same as in the previous section
with the only difference that we have to construct the new vector
$\vec{b}$ and subtracts it as was indicated in Eq. (\ref{methodST}),
i.e. we can use the results given above for the analytical inverse
$\mathbf{D}_{2}^{-1}$ and for the product
$\mathbf{E}=\mathbf{D}_{2}^{-1}\mathbf{D}_{1}$.

\subsection{Plane Waves with Constant Amplitude}
\begin{figure}[h!]
\begin{center}
\includegraphics[width=4cm,angle=270]{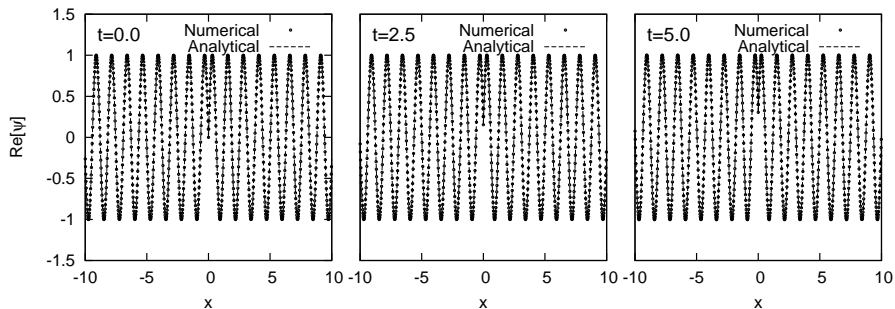}
\vspace{0.5cm}
\caption{Re$[\psi]$ as a function of distance $x$, for the analytical (dashed line) and
numerical (dots) solutions to the Schr\"{o}dinger equation with source
term. The initial conditions are $x_0=0$, $t_{0}=0$, $\alpha_1=12$
$\alpha_2=13$, $S_0=5$, $\omega=(p_0-g_{2})/g_{1}$ and $p_0=5$. The
left and right boundaries of the domain are $x_{\rm min} = -10$ and
$x_{\rm max} = 10$ respectively. The size of every lattice in the grid
is $\Delta x = 0.02$ with $N=1000$ discrete lattices in all the
spatial domain and the time step is $\Delta t = 0.001$. With the
given parameters the matrix elements for (\ref{d1n}) are calculated from Eq. (\ref{param_tid}). The given times are for
the variable $t=t_{0}+ n \Delta t$.} \label{compST}
\end{center}
\end{figure}

Here we consider the case $V(x)=0$ and $S(t)=S_0$ in
which the exact solution was given as
\begin{equation}\label{STctte}
\psi(x,t)=\frac{S_0}{ik}\exp(ik|x|)\exp(-i\omega t).
\end{equation}
In Fig. \ref{compST} we compare the exact Eq. (\ref{STctte}) and the
numerical results for this case. The agreement between the numerical and exact result suggest that the method is sufficiently accurate and stable. It should be noted the excellent behavior of the Absorbing Boundary Conditions, which, regardless that the source is filling the numerical region, the numerical evolution simulates an open domain.

\subsection{Plane Waves with time--dependent Amplitude}
\begin{figure}[h!]
\begin{center}
\includegraphics[scale=0.4,angle=270]{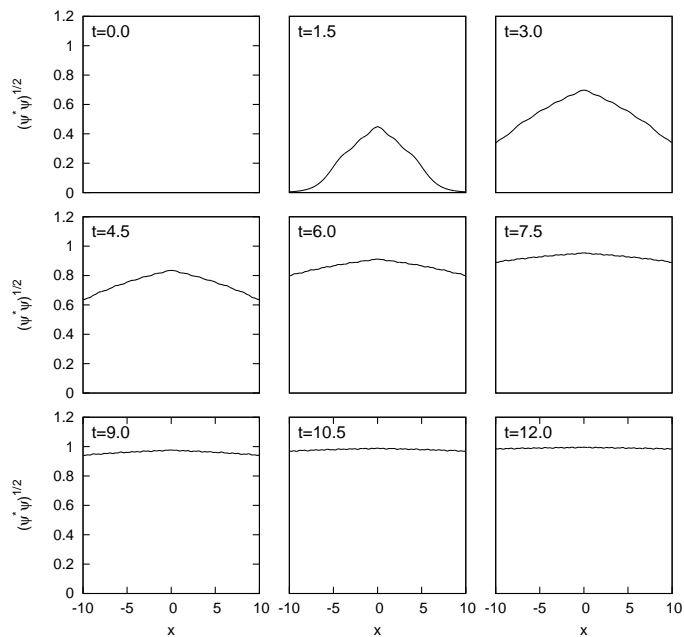}
\caption{$|\psi|$ as a function of distance $x$, for the numerical
solution to the  equation with time--dependent source
term.} \label{denpsi}
\end{center}
\end{figure}

The case of plane waves with time--dependent amplitude is of physical interest, because the idea
of an initially empty waveguide that is gradually filled with matter
waves corresponds to the experimental realization of a reservoir
located at $x=x_0$. For propagation times $t\gg \Delta t$, the calculation converges toward a
flat density that corresponds to the stationary plane waves at the
source amplitude $S=S_0$. The time evolution of the probability density during the increase of the source
amplitude is displayed in Fig. \ref{denpsi}.

\begin{figure}[h!]
\begin{center}
\includegraphics[scale=0.3,angle=270]{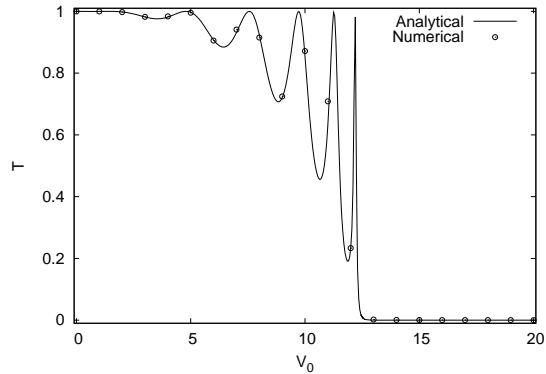}
\caption{Comparison of the analytical and numerical results for the transmission probability $T$ of a finite potential barrier, as a function of the potential height $V_{0}$. The same parameters as in Fig. \ref{compST}} \label{trapb}
\end{center}
\end{figure}

The transmission through a potential barrier has been a model of great relevance from a pedagogical point of view, as discussed in almost every textbook on quantum mechanics. The exact solution to this problem is usually obtained by assuming that a plane wave impinges on the barrier from the left. However, the comparison of the analytical result for the transmission and/or reflection coefficients with the numerical one is not an easy task, because the usual calculation in terms of gaussian wave-packets gives us an average of the analytically calculated transmission coefficients. This fact can be understood taking in to account that a Gaussian wave packet can be viewed as a superposition of plane waves with different momentum. Then, the correct way to compare the transmission and/or reflection coefficients in this case is to solve the numerical problem with a source term emitting plane-waves. In Fig. \ref{trapb}, we compare the analytical transmission coefficients for the case of a finite potential barrier with the predicted by our numerical method with time--dependent amplitude. Also in this case we find an excellent agreement with the exact results.

\section{Summary}
We have described a numerical integration method for the time--dependent Schr\"{o}dinger equation with an without source terms. In particular, we consider the case of scattering systems, in which the Dirichlet boundary conditions produces undesired reflections. To solve this problem we have introduced absorbing boundary conditions (ABC) into this integration scheme. The numerical integration was done using the Crank-Nicholson method together with a Cayley's form for the finite-difference representation of evolution operator which produces an stable, unitary, and second-order accurate in space and time method. On replacing the Hamiltonian by its finite-difference approximation, the problem reduces to a complex tridiagonal system. We have simplified the numerical scheme by inverting analytically the matrices by means of the Usmani's formula for Jacobian matrices. The analytical inversion of the matrices decrease the computational effort and let us fully and easily implement the method into Mathematica or even into any scientific computational software. This numerical method can be used for arbitrary potential shapes even in the presence of the source term, which does not limit its use to elementary applications required in teaching quantum mechanics. The formalism discussed here may be extended in a straightforward way to time--dependent potentials, in which the matrices vary in every time step. Finally, we have compared the results of our modified numerical method with the analytical solution for the transmission probability of a finite potential barrier, where we find an excellent level of agreement.

\section*{Acknowledgments}
We enjoyed fruitful discussions with K. Rapedius, M. Hartung, L. A. Pachon, T. Dittrich, K. Richter, and C. Viviescas. Financial support from Volkswagen Foundation (grant I/78235), Universidad Nacional de Colombia in the program Becas para Estudiantes Sobresalientes de Posgrado, Colciencias, and the ALECOL program of the German Academic Exchange Service DAAD is gratefully acknowledged. We thank for the hospitality extended to us by the MPI for the Physics of Complex Systems, Dresden, University of Technology, Kaiserslautern, and University of Regensburg, where part of this work has been carried out.

\end{document}